\documentclass{jps-cp}
\usepackage{txfonts} %

\title{About Toroidal Model of Leptons in Space-Time Film Theory}

\author{Alexander A. \textsc{Chernitskii}$^{1,2}$}

\inst{$^{1}$Department  of Mathematics, St. Petersburg State Chemical and Pharmaceutical University,\\ Prof. Popov str. 14, St. Petersburg, 197376, Russia \\
$^{2}$A.~Friedmann Laboratory for Theoretical Physics, St.-Petersburg, Russia}

\email{alexander.chernitskii@pharminnotech.com}

\recdate{March 10, 2022}

\abst{Nonlinear field model of extremal space-time film is considered.
Its space-localized solutions in toroidal coordinates with periodic dependence on time are investigated.
An accordance between these soliton solutions and leptons is considered.
}

\kword{space-time film, toroidal soliton-particle, lepton}

\usepackage{fancyhdr}
\fancypagestyle{firststyle} 
{
\fancyhead[L]{{\it Proc. 24th International Spin Symposium (SPIN2021)}\\
JPS Conf. Proc. {\bf 37}, 020608 (2022)\\
htps://doi.org/10.7566/JPSCP.37.020608
}
\fancyhead[R]{}
\fancyfoot[C]{{\bf 020608-\thepage}}
\fancyfoot[R]{\copyright 2022 The Author}
}

\def\newblock{}
\def\fZ#1{$#1$}

\def\p{\partial}

\def\stTD#1#2{\hbox to 0em{\mathsurround=0em $\stackrel{#1}{\makebox[0pt]{} #2}$\hss} \phantom{#2}}\def\stscript#1#2{\hbox to 0em{\mathsurround=0em ${\scriptstyle\stackrel{#1}{\makebox[0pt]{} #2}}$\hss} \phantom{#2}}\def\stscriptscript#1#2{\hbox to 0em{\mathsurround=0em ${\scriptscriptstyle\stackrel{#1}{\makebox[0pt]{} #2}}$\hss} \phantom{#2}}
\def\comb#1#2#3{{\mathsurround 0pt\hbox to 0pt {\hspace*{#3}\raisebox{#2}{${#1}$}\hss}}}
\def\combs#1#2#3{{\mathsurround 0pt\hbox to 0pt {\hspace*{#3}\raisebox{#2}{${\scriptstyle #1}$}\hss}}}
\def\combss#1#2#3{{\mathsurround 0pt\hbox to 0pt {\hspace*{#3}\raisebox{#2}{${\scriptscriptstyle #1}$}\hss}}}
\def\e#1{\mathrm{e}^{#1}}

\def\csch{\operatorname{csch}}

\def\cep{\bar{e}}

\def\tAct{\tilde{\mathcal{A}}}
\def\Act{\mathcal{A}}

\def\df{\mathrm{d}}

\def\lambdaC{\lambda_{\!\!\!C}}

\def\CLF{\mathcal{K}}
\def\CLF1{\mathcal{K}_1}
\def\CLF2{\mathcal{K}_2}

\def\metrEff{\mathchoice{\combs{\sim}{1ex}{0.2ex}\mathfrak{m}}{\combs{\sim}{1ex}{0.2ex}\mathfrak{m}}{\combss{\sim}{0.66ex}{0.05ex}\mathfrak{m}}{}{}}
\def\metrEffm1{\check{\metrEff}}

\def\metr{\mathfrak{m}}

\def\Vol{\overline{V}}
\def\dVol{\df\mspace{-2mu}\Vol}
\def\Vols{V}

\def\dVols{\df\mspace{-2mu}\Vols}

\def\p{\partial}

\def\OOO#1#2{\mathcal{O}\!\left(#1\right)_{#2}}

\def\epar{\varepsilon}

\def\LF{\mathchoice{\combs{-}{0.3ex}{0.3ex}\mathcal{L}}{\combs{-}{0.3ex}{0.3ex}\mathcal{L}}{\combss{-}{0.25ex}{0.3ex}\mathcal{L}}{}{}}

\def\bje{{\mathsurround 0pt\lower.0ex\hbox{${\scriptscriptstyle \mathbf{e}}$}\mspace{-3.4mu}\mathbf{j}}}
\def\je{{\mathsurround 0pt\lower.0ex\hbox{${\scriptscriptstyle e}$}\mspace{-4.5mu}j}}
\def\bjm{{\mathsurround 0pt\lower.0ex\hbox{${\scriptscriptstyle \mathbf{m}}$}\mspace{-5.6mu}\mathbf{j}}}
\def\jm{{\mathsurround 0pt\lower.0ex\hbox{${\scriptscriptstyle m}$}\mspace{-7.0mu}j}}

\def\cE{\mathcal{E}}

\def\cM{\mathcal{J}}

\def\ffun{\Phi}

\def\bffun{\underline{\Phi}}
\def\tffun{\bar{\Phi}}

\def\ffun{\Phi}

\def\ffind{\Upsilon}

\def\ccc{\mathchoice{\combs{-}{0.07ex}{-0.2ex}{\mathrm{c}}}{\combs{-}{0.07ex}{-0.2ex}{\mathrm{c}}}{\combss{-}{0.01ex}{-0.25ex}{\mathrm{c}}}{}{}}
\def\xxx{\chi}
\def\balpha{\mathchoice{\combs{-}{0.04ex}{-0.1ex}{\alpha}}{\combs{-}{0.045ex}{-0.1ex}{\alpha}}{\combss{-}{0.03ex}{-0.15ex}{\alpha}}{}{}}

\def\bPlank{\hbar}
\def\bbPlank{\mathchoice{\combs{=}{0.07ex}{-0.1ex}{h}}{\combs{=}{0.07ex}{-0.1ex}{h}}{\combss{=}{0.01ex}{-0.15ex}{h}}{}{}}

\def\eqdef{\doteqdot}

\def\brhoo{\breve{\rhoo}}
\def\brrho{\bar{\bar{\rho}}}
\def\brekappa{\breve{\kappa}}
\def\brebkappa{\breve{\bkappa}}

\def\bkappa{\bar{\kappa}}

\def\mtheta{\bar{\theta}}
\def\cmtheta{\bar{\bar{\theta}}}
\def\vups{\upsilon}
\def\rhoo{\rho_{\!\circ}}

\def\romega{\tilde{\omega}}

\def\bmclK{\bar{\mathcal{K}}}

\def\tAMV{\tilde{\mathbb{J}}}
\def\Energy{\mathbb{E}}

\def\AMV{\mathbb{J}}

\begin{document}

\maketitle

\thispagestyle{firststyle}

\section{Introduction}
\label{introd}
In this paper, we consider toroidal particle-like solutions or particles-solitons of the space-time film equation \cite{Chernitskii2018a} which can be associated with elementary particles, in particular, with leptons.

Toroidal solitons was considered by author earlier in the scope of nonlinear electrodynamics \cite{Chernitskii2012be}. Some results of the investigation for toroidal solitons of space-time film were published in  \cite{Chernitskii2020a}.

\section{Field model and equations}
\label{fmodeqs}
The model action has the following form \cite{Chernitskii2018a}:
\begin{equation}
\label{382010071}
\Act  = \int\limits_{\Vol}\LF\;\dVol
\;,\quad
\LF  \eqdef
\sqrt{\left|1 + \xxx^{2}\,\metr^{\mu\nu}\,\frac{\p \ffun}{\p x^{\mu}}\,\frac{\p \ffun}{\p x^{\nu}}\right| }
\end{equation}
where $\dVol \eqdef \sqrt{|\metr|}\;\left(\df x\right)^{4}$ is the four-dimensional volume element, $\metr \eqdef \det(\metr_{\mu\nu})$,
$\ffun$ is the scalar field function,
$\metr_{\mu\nu}$,
$\metr^{\mu\nu}$ are components of metric tensor for an arbitrary coordinate system in flat space-time.
The Greek indexes take the values $\{0,1,2,3\}$. $x^0\eqdef \ccc\,t$ is time coordinate, where $t$ is time and $\ccc$ is the speed of light.
Here we consider both signatures of metric $\{+,-,-,-\}$ and $\{-,+,+,+\}$, that is
 $\metr^{00} = 1$ and $\metr^{00} = -1$.

The stationary condition for the action (\ref{382010071}) gives the following equation:
\begin{equation}
\label{45750360}
  \frac{1}{\sqrt{|\metr|}}\,\frac{\p }{\p x^{\mu}} \frac{\sqrt{|\metr|}\,\metr^{\mu\nu}}{\LF}\,\frac{\p\ffun}{\p x^{\nu}} = 0
\;,
\end{equation}

Let us introduce the coordinate system which can be called rational toroidal one.
This system is obtained from the usual toroidal one $\{x^{0},\kappa,\upsilon,\varphi\}$ by the following change of variable $\kappa$ ($0\leqslant\kappa\leqslant\infty$):
$
\bkappa  = \e{\kappa} - 1
$.
We have for the new coordinate system the following useful relations:
\begin{equation}
\label{624492681}
\cosh\kappa  = \frac{2 + \bkappa\left(\bkappa + 2\right)}{2 \left(\bkappa + 1\right)}\;,
\quad
\sinh\kappa  = \frac{\bkappa\left(\bkappa + 2\right)}{2 \left(\bkappa + 1\right)}
\;.
\end{equation}

\begin{subequations}\label{739075901}
The diagonal components of the metric tensor for this coordinate system have the form:
\begin{align}
\label{539540481}
 &\metr_{00}=\pm 1\;,\quad
\metr_{\bkappa\bkappa} = \mp\frac{\rho_\circ^2}{4\,\bmclK^2}\;,\quad
\metr_{\upsilon\upsilon} = \mp\frac{\left(\bkappa + 1 \right)^2 \rho_\circ^2}{4\,\bmclK^2}\;,\quad
\metr_{\varphi\varphi} = \mp\frac{\bkappa^{2}\left(\bkappa + 2\right)^2 \rho_\circ^2 }{16\,\bmclK^2}\;,
\\
\label{62387657}
& \bmclK\eqdef \frac{1}{4}\,\bigl(2 + \bkappa \left(\bkappa + 2\right) - 2 \left(\bkappa +1\right) \cos\upsilon\bigr)
= (\bkappa+1)\, \sin^2\!\left(\frac{\vups }{2}\right) + \frac{\bkappa^2}{4}
\;.
\end{align}
\end{subequations}

Let us introduce also the rational toroidal coordinate system with rotation $\{\mtheta,\cmtheta,\bkappa,\upsilon\}$,
which is obtained by the following change of variables:
\begin{equation}
\label{592268531}
 \mtheta = \varphi - \romega\,x^{0}
 \;,\quad
 \cmtheta = \varphi + \romega\,x^{0}
\;.
\end{equation}
The coordinates $\mtheta$ and $\cmtheta$ can be called right and left phase coordinates accordingly, and the parameter $\romega$
is an angular velocity.

In this work, we take the following relation between the angular velocity and the radius of a singular ring for the coordinate system:
\begin{equation}
\label{794493791}
\romega  =
\frac{1}{\rhoo}
\;.
\end{equation}

It is evident that the function depending on the three variables only $\{\mtheta,\bkappa,\upsilon\}$ or $\{\cmtheta,\bkappa,\upsilon\}$
is right- or left-rotating field configuration about $x^{3}$ axis. Because of condition (\ref{794493791}),
the phase velocities for the appropriated ring waves equal the speed of light on the singular ring.

\begin{subequations}\label{723178271}
The energy density $\cE$ and z-component of the angular momentum density $\cM_{z}$ have the following form in
the coordinate system with the rotation:
\begin{align}
\label{72338185}
\cE &=
\frac{1}{4\pi}\,\left(\frac{1}{\rhoo^2\,\LF}\left(\frac{\p\ffun}{\p \mtheta} - \frac{\p\ffun}{\p \cmtheta}\right)^{\!2} \pm\frac{1}{\xxx^{2}}\,\bigl(1 - \LF\bigr)\right)
\;,\\
\label{72355080}
\cM_{z} &= \frac{1}{4\pi\,\rhoo\,\LF}\left(\left(\frac{\p\ffun}{\p \mtheta}\right)^{\!2} -
\left(\frac{\p\ffun}{\p \cmtheta}\right)^{\!2}\right)
\;,
\end{align}
\end{subequations}

Let us introduce the following dimensionless parameter for a solution:
\begin{equation}
\label{751143031}
\epar  \eqdef \pm \frac{\cep^2\,\xxx^2}{\rhoo^{4}}
\;,
\end{equation}
where the sign of $\epar$ coincides with the sign of $\metr^{00}$.

Here we will consider a rotated field configuration depending on the three variables $\{\mtheta,\bkappa,\upsilon\}$. For such a solution we have the following notable expressions:
\begin{equation}
\label{776342691}
\cE   =
\romega\left(\cM_{z} + \frac{\cep^2}{4\pi\,\rhoo^3\,\epar}\,\bigl(1 - \LF\bigr)\right)
\;,\quad
\cM_{z} = \frac{1}{4\pi\,\rhoo\,\LF}\left(\frac{\p\ffun}{\p \mtheta}\right)^{\!2}
\end{equation}

Energy and angular momentum of a soliton-particle are
\begin{equation}
\label{576969441}
 \Energy \doteqdot
\int\limits_{\Vols}\cE\,\dVols
\;,\quad
\AMV \doteqdot \int\limits_{\Vols}\cM_{z}\,\dVols
\;,
\end{equation}
where $\Vols$ is all three-dimensional space.

We will search for a solution of the model equation for the dimensionless function $\tffun$ such that
\begin{equation}
\label{567210181}
\ffun  = \frac{\cep}{\rhoo}\,\sqrt{\bmclK}\;
\tffun
\;,
\end{equation}
Here the factor can be represented in spherical coordinates as
\begin{equation}
\label{369227981}
\frac{\cep}{\rhoo}\,\sqrt{\bmclK}  = \frac{\cep}{r} +\frac{\cep\,\rhoo\,\sin\vartheta}{r^2} + \frac{\cep\,\rhoo^2 \left(1 - 3\, \cos(2\,\vartheta )\right)}{4\, r^3}+
\OOO{r^{-4}}{r\to \infty}
\;.
\end{equation}

\section{Lepton as toroidal soliton-particle}
\label{lepttorsol}
To have a real representation of elementary particles with the solitons of space-time film, we must obtain the observable discrete mass spectrum of the particles. But the field model under consideration has a scale invariance which forbids the
discrete spectrum of the energy for solutions. This criticism of unified field theories has been known for a long time and the appropriated response was given by A. Einstein \cite{Einstein1953aE}.
The essence of this response is that the scale invariance of a soliton solution is broken for a multy-soliton solution. In other words, the interaction between solitons-particles breaks the scale invariance for separate soliton-particles.

Let us consider the scale invariance by the example of the simplest spheroidal soliton-particle or spheron \cite{Chernitskii2017a}. An application of the scale transformation with the parameter $a$ for this spheron solution is equivalent to
the change of value of charge $\cep\to a^2 \cep$. That is the free parameter $\cep$ of the spheron solution is connected with the scale invariance. Thus if we fix the value of charge $\cep$ than we fix the scale of the soliton and all other its parameters.
But the problem of quantization of charge can not be solved by the consideration of a separate soliton. It must be considered taking into account the interaction with other soliton solutions.
Here we take that the value of charge equals the elementary electrical charge.

We will express all dimensional quantities by the dimensions of electrical potential and length that is by volt (V) and metre (m).
We can use the two basic units only because we consider dependencies from time coordinate $x^0\eqdef \ccc\,t$ instead of time $t$.
We call this system of units VM.
Let us compare the systems of units SI and VM.
\begin{equation}
\label{739508881}
\frac{q}{4\pi\,\epsilon_0\,r}  = \frac{\cep}{r}\quad\Longrightarrow\quad \cep = \frac{q}{4\pi\,\epsilon_0}\approx
1.440\cdot 10^{-9}\,\text{V}\cdot\text{m}
\;,
\end{equation}
where $q \approx 1.602\cdot 10^{-19}\;\text{C}$ is elementary electrical charge in SI systems of units, $\epsilon_0\approx 8.854\cdot 10^{-12}\;\text{F}/\text{m} $ is electric constant.

In the VM system of units, energy $\Energy$ and angular momentum $\AMV$
have the dimensions $\text{V}^2\,\text{m}$ and $\text{V}^2\,\text{m}^2$ accordingly.
\begin{subequations}\label{482713801}
\begin{align}
\label{482823311}
1\,\text{eV} &\approx 1.440\cdot 10^{-9}\,\text{V}^2\cdot\text{m}
\;,\\
\label{482823312}
\bbPlank \eqdef \ccc\,\bPlank \approx 1.973\cdot 10^{-7}\,\text{eV}\cdot\text{m}
&\approx
2.842\cdot 10^{-16}\,\text{V}^2\cdot\text{m}^2
\;.
\end{align}
where $\ccc$ is the speed of light, $\bPlank$ is usual Planck's constant.
\end{subequations}
As we see we have the right value of the fine structure constant $\balpha$:
\begin{equation}
\label{577592591}
\balpha^{-1} = \frac{\bbPlank}{\cep^2} \approx 137
\;.
\end{equation}

In the VM system, the Compton wave-length is given by the following empirical formula:
\begin{equation}
\label{612152841}
\lambdaC  = \frac{2\pi\,\bbPlank}{\Energy}
\;.
\end{equation}

We suppose that the Compton wave-length of a particle connects uniquely  with the fundamental frequency $\nu$ of an oscillatory part of the appropriate soliton. This connection is given in the VM system of units by formula
\begin{equation}
\label{764055291}
\nu = \frac{\omega}{2\pi} = \lambdaC^{-1}
\;.
\end{equation}

Taking into account (\ref{612152841}) and (\ref{764055291}), we have the empirical connection:
\begin{equation}
\label{801446051}
 \Energy = \bbPlank\,\omega
\;.
\end{equation}

We suppose that the whole number of the Compton wavelengths is kept within the singular ring.
Then taking into account (\ref{764055291}) and (\ref{794493791}) we have
\begin{equation}
\label{778538521}
2\pi\,\rhoo  = n\,\lambdaC\quad\Longrightarrow\quad
\rhoo = \frac{n}{\omega}
\quad\Longrightarrow\quad
\omega = n\,\romega
\;.
\end{equation}

The introduced quantum number $n$ in (\ref{778538521}) can define different solitons-particles.

Considering the lepton family, let us write the experimental value of their angular momentum or spin:
\begin{equation}
\label{835633201}
\AMV  = \frac{\bbPlank}{2}
\;.
\end{equation}

Taking into account (\ref{723178271}) -- (\ref{576969441}) and (\ref{778538521}),
we deduce that the empirical relations (\ref{801446051}) and (\ref{835633201}) need the following relations for a rotating soliton-particle:
\begin{subequations}\label{825005531}
\begin{alignat}{3}
\label{825053731}
\tAMV\eqdef\frac{1}{\cep^2}\,\AMV\, &=\, \int\limits_{\Vols}
\frac{1}{4\pi\,\rhoo^{3}\,\LF}\left(\frac{\p\bffun}{\p \mtheta}\right)^{\!2}\dVols  \;&=&\;\, \frac{\balpha^{-1}}{2}
\;,\\
\label{825053732}
\tAct\eqdef\frac{1}{\cep^2}\,\Bigl(\rhoo\,\Energy - \AMV\Bigr)\, &=\,  \int\limits_{\Vols}
\frac{1}{4\pi\,\rhoo^3\,\epar}\,\bigl(1 - \LF\bigr)\,\dVols  \;&=&\;\; \balpha^{-1}\left(n - \frac{1}{2}\right)
\;,
\end{alignat}
where \fZ{\tAMV} and \fZ{\tAct} are dimensionless quantities.
\end{subequations}

Using the designations (\ref{825005531}) let us write the following expression for energy of a rotating soliton-particle:
\begin{equation}
\label{366134701}
\Energy  = \frac{\cep^2}{\rhoo}\left(\tAMV\ + \tAct\right) = \frac{n\,\cep^2}{\balpha\,\rhoo}
\;.
\end{equation}

\def\qL{\bar{\bar{q}}}
It is reasonable that a desired toroidal solution near the singular ring is similar to a solution with some axial symmetry.
In particular a static part of the toroidal solution can be the circinate charged string. The exact solution in the form of charged string was obtained in \cite{Chernitskii2018a}).
This solution has the following characteristic radial dimension \fZ{\brrho} and radial component of electrical field \fZ{\ffind_{\rho}}:
\begin{equation}
\label{561489931}
\brrho\eqdef \left|2\,\qL\,\xxx\right|\,,\qquad \ffind_{\rho} =-\frac{2\,\qL}{\rho}
\;,
\end{equation}
where \fZ{\qL} is a linear charge density, \fZ{\rho} is the cylindrical radial coordinate.
Recent investigation connected with gravitational interaction in the space-time film theory \cite{Chernitskii2021b} denotes the negative dimensionless parameter \fZ{\epar < 0}.
In this case the surface \fZ{\rho = \brrho} for the string solution is singular.

We can suppose that the appropriate singular surface for the circinate charged string is the toroidal surface
with its minor radius equal \fZ{\brrho}. For this case we have
\begin{equation}
\label{261297401}
\brhoo  =  \rhoo\,\coth\brekappa = \rhoo\left(\frac{1}{\brebkappa} + \frac{1 + \brebkappa}{2 + \brebkappa}\right)\,,\quad
\brrho = \rhoo\,\csch\brekappa = \rhoo\left(\frac{1}{\brebkappa} + \frac{1}{2 + \brebkappa}\right)
\;,
\end{equation}
where \fZ{\brekappa} and \fZ{\brebkappa} are values of coordinates \fZ{\kappa} and \fZ{\bkappa} which are
appropriated to toroid with minor radius \fZ{\brrho} and major radius \fZ{\brhoo}.
The full charge of the circinate string must be equal to the elementary electrical charge:
\begin{equation}
\label{591238551}
 2\,\pi\,\brhoo\,\qL = \cep
\;.
\end{equation}

Taking into account (\ref{561489931}), (\ref{261297401}), (\ref{591238551})
and
(\ref{751143031}) we obtain the following expresion:
\begin{equation}
\label{334837871}
\brebkappa  = \frac{\pi + \sqrt{\pi^2 + 4 \left| \epar \right|}+\sqrt{2 \pi } \sqrt{\pi + \sqrt{\pi ^2 + 4 \left| \epar \right| }}}{2 \sqrt{\left| \epar \right| }} - 1
\;.
\end{equation}

For the case \fZ{\brrho \ll \rhoo} we have
\begin{equation}
\label{340108261}
\brhoo  \approx \rhoo\quad\Longrightarrow \quad \brebkappa \approx \frac{2\,\pi}{\sqrt{\left| \epar \right| }}
\;,
\end{equation}
that conforms with the formula (\ref{334837871}) for the case \fZ{\left| \epar \right|\ll 1}.

But we must take into account that the simple turning of the cylindrical solution does not give an exact solution.
Moreover we consider a solution with dependence on time. Thus an appropriate singular surface may be different from
a toroidal one \fZ{\bkappa = \brebkappa}.

We search the solution for the function \fZ{\tffun = \tffun(\bkappa,\mtheta,\upsilon)} (see (\ref{567210181})) that is a rotating solution.
The approximate solution is represented in the form of partial sum for a formal power series in $\bkappa$:
\begin{equation}
\label{375684771}
 \tffun = \sum\limits_{i=0}^{N} \tffun_{i}\,\bkappa^{i}
\;,
\end{equation}
where the coefficients $\tffun_{i}$ are periodical functions in two coordinates $\tffun_{i} = \tffun_{i}(\mtheta,\upsilon)$.

We substitute the representation (\ref{375684771}) to a model equation.
Then the coefficients for different powers of $\bkappa$ equate with zero.
The obtained equations are linear differential equation of the second order on the variable \fZ{\mtheta}.
And these equations does not contain derivatives on the variable \fZ{\upsilon}.

Each such equation can be solved by decomposition in harmonic components on the variable \fZ{\mtheta}.
The appropriate solution for the function \fZ{\tffun_{i}} contains arbitrary functions on the variable \fZ{\upsilon}. These arbitrary functions and their derivatives are contained in equations for the function \fZ{\tffun_{i+1}} and
the following functions. Existence conditions for regular solutions can specify the arbitrary functions.

\section{Conclusions}
\label{concl}
Thus we have considered the problem for finding the toroidal soliton solutions of the space-time film equation. These solutions can be correlated with leptons.

We have introduced the rational toroidal coordinate system which is more appropriate for the problem.

The criterions of association of the soliton solutions with leptons are obtained.

The method for finding the solution in the form of formal power series in the variable \fZ{\bkappa} is proposed.
The approximate solution with terms up to \fZ{\bkappa^{3}} is obtained.

\end{document}